\documentclass[aps,amsmath,twocolumn,amssymb,titlepage,superscriptaddress,10pt]{revtex4-1}
\usepackage[T1]{fontenc}
\usepackage[utf8]{inputenc}
\usepackage{amsmath}
\usepackage{braket}
\usepackage{amsfonts}
\usepackage{comment}
\usepackage{graphicx}
\usepackage[breaklinks,colorlinks,bookmarks=false,citecolor=blue,linkcolor=red,urlcolor=blue]{hyperref}
\usepackage[capitalize]{cleveref}
\usepackage{amssymb}
\usepackage{amsmath}
\usepackage{mathtools}
\usepackage{todonotes}

\begin{document}
\title{Phases of the spin-1/2 Heisenberg antiferromagnet on the diamond-decorated\\ square lattice in a magnetic field}
\author{Nils Caci}
\affiliation{Institute for Theoretical Solid State Physics, JARA FIT, and JARA CSD, RWTH Aachen University, 52056 Aachen, Germany}
\author{Katar\'ina Karl'ov\'a}
\affiliation{Department of Theoretical Physics and Astrophysics, Faculty of Science, P.~J. \v{S}af\'arik University, Park Angelinum 9, 04001 Ko\v{s}ice, Slovakia}
\author{Taras Verkholyak}
\affiliation{Institute for Condensed Matter Physics, National Academy of
Sciences of Ukraine, Svientsitskii Street 1, 790 11, L'viv, Ukraine}
\author{Jozef Stre\v{c}ka}
\affiliation{Department of Theoretical Physics and Astrophysics, Faculty of Science, P.~J. \v{S}af\'arik University, Park Angelinum 9, 04001 Ko\v{s}ice, Slovakia}
\author{Stefan Wessel}
\affiliation{Institute for Theoretical Solid State Physics, JARA FIT, and JARA CSD, RWTH Aachen University, 52056 Aachen, Germany}
\author{Andreas Honecker}
\affiliation{Laboratoire de Physique Théorique et Modélisation, CNRS UMR 8089, \\ CY Cergy Paris Université, 95000 Cergy-Pontoise, France}

\begin{abstract}
The spin-1/2 Heisenberg antiferromagnet on the frustrated diamond-decorated square lattice is known to  feature various zero-field ground-state phases, consisting of extended monomer-dimer and  dimer-tetramer ground states  as well as a ferrimagnetic regime.  Using a combination of analytical arguments,  density matrix renormalization group (DMRG), exact diagonalization, as well as  sign-problem-free quantum Monte Carlo (QMC) calculations, we investigate the properties of this system and the related Lieb lattice in the presence of a finite magnetic field, addressing both the ground-state phase diagram as well as several thermodynamic properties. In addition to the zero-field ground states, we find at high magnetic field a spin-canted phase with a continuously rising magnetization for increasing magnetic field strength, as well as the fully polarized paramagnetic phase. At intermediate field strength, we identify a first-order quantum phase transition line between the ferrimagnetic  and the monomer-dimer regime. This first-order line  extends  to finite temperatures, terminating in a line of critical points that belong to the universality class of the two-dimensional Ising model. 
\end{abstract}

\date{March 1, 2023}

\maketitle

\section{Introduction}
The study of strongly frustrated quantum magnets is a central topic in contemporary
condensed matter research. Indeed, magnetic frustration, introduced, e.g.,
by competing antiferromagnetic exchange couplings, can lead to the stabilization of
non-classical ground states in quantum magnets \cite{Richter2004,Balents10,HFMbook,DIEPbook}.
In most cases, these non-magnetic states are characterized by the formation of strong local singlets among small sub-clusters of
spins, as well as the emergence of an extensive ground-state entropy. In the most
favorable case, it is possible to obtain  exact analytical expressions for the
ground-state properties, such as for the Shastry-Sutherland model in the regime
of strong dimer coupling
\cite{Shastry1981,Albrecht1996,Miyahara1999,MiUeda03}.
In this system, quantum spin 
degrees of freedom are arranged on a two-dimensional lattice in an orthogonal
manner to form a frustrated array of coupled spin dimers. For strong intra-dimer
coupling (as compared to the inter-dimer coupling), an exact product state of
dimer singlets forms the system's ground state. Later, it was furthermore found
that the spin-1/2 version of this quantum spin model finds an almost perfect realization
in the copper-based compound SrCu$_2$(BO$_3$)$_2$
\cite{Kageyama1999,Miyahara1999}. This system has since then been studied
extensively with respect to both the ground state
and thermal properties \cite{Kageyama1999,Kageyama2000,Kageyama2000b,Lemmens00,Gaulin2004,Zayed2014,Zayed2017,Guo2020,Jimenez2021}
as well as its rich physics in the additional presence of a  magnetic field,
notably various plateaux in its magnetization curve
\cite{Kageyama1999,Onizuka2000,Kodama2002,Sebastian2008,Takigawa2011,Takigawa2012,Jaime2012,Matsuda2013,Haravifard2016}.

\begin{figure}[t!]
    \centering
    \includegraphics{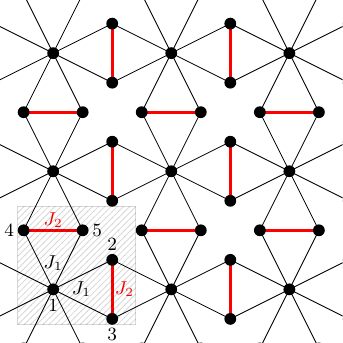}
    \caption{Illustration of the diamond-decorated square lattice, with a unit cell indicated (dashed square), along with the labeling of the five different sites (circles) within the unit cell and the two different exchange couplings $J_1$ (thin black lines) and $J_2$ (thick red lines).}
    \label{fig:lattice}
\vspace*{-3mm}
\end{figure}

The dimerized nature of the low-energy states in the Shastry-Sutherland model not only gives rise to interesting physics, but is actually also favorable for a numerical treatment. Indeed, the Shastry-Sutherland model  is not only a showcase for tensor-network approaches \cite{Takigawa2012,CorbozMila13,Corboz2015,Wessel2018,Wietek19,Jimenez2021}, but it also allows one to use efficient Quantum Monte Carlo (QMC) simulations throughout a large part of the dimer phase \cite{Wessel2018,Wietek19,Honecker2022}.
Remarkably, the latter extends to a generalized version of the Shastry-Sutherland model \cite{Weihong99,PhysRevLett.84.1808} where in a certain limit, that is equivalent to a fully frustrated bilayer model \cite{Hida92,Sandvik94,Sommer2001,Wang06},
the QMC sign problem disappears completely.
The fully frustrated bilayer model thus becomes accessible to detailed investigations at finite temperature via QMC
simulations \cite{Alet16,NgYang17,Stapmanns2018}. In fact, the identification of a first-order line that terminates at a finite-temperature
critical point \cite{Stapmanns2018} in the fully frustrated bilayer model was an important guiding element to identify similar physics
in the Shastry-Sutherland model and ultimately SrCu$_2$(BO$_3$)$_2$ \cite{Jimenez2021}.
Note, furthermore, that the low-energy high-field region of the fully frustrated bilayer model permits a mapping  to a classical lattice gas, thus allowing for a rigorous treatment of its low-energy thermodynamics, including a finite-temperature ordering transition \cite{RDK06,DRHS07}.

\begin{figure}[t!]
    \centering
    \includegraphics{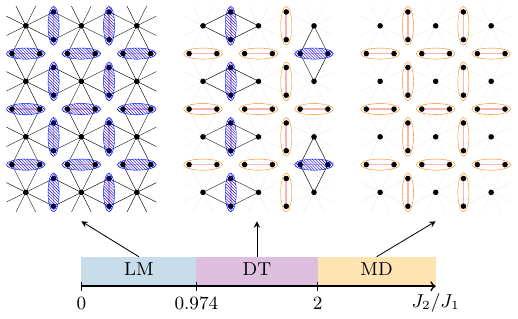}
    \caption{Ground-state phase diagram of the spin-1/2 Heisenberg antiferromagnet on the diamond-decorated square lattice in zero magnetic  field as obtained from Ref.~\cite{Hirose2017} containing the Lieb-Mattis (LM), dimer-tetramer (DT) and the monomer-dimer (MD) phase. In the illustration of the different ground states, blue (orange) ovals denote spin triplet (singlet) states on the dimers. A tetramer singlet of the DT phase is illustrated by a rhombus.}
    \label{fig:h0_phasediag}
\vspace*{-3mm}
\end{figure}

Another highly frustrated two-dimensional quantum spin system of coupled orthogonal spin dimers
is the 
Heisenberg antiferromagnet on the diamond-decorated square lattice, shown in Fig.~\ref{fig:lattice}.
This model contains, in addition to the dimers (along the $J_2$ bonds), a further set of spins
that are coupled to other (dimer) spins only by the $J_1$ bonds. In the large $J_2$-limit, for $J_1/J_2\rightarrow 0$, the spins coupled solely through 
the $J_1$ bonds thus lack a partner spin to form a singlet, and we therefore refer to these spins as monomer spins.
Hirose \textit{et al.}\ have performed a detailed investigation of its zero-field ground-state properties
\cite{Morita2016,Hirose2016,Hirose2017,Hirose2018,hirose}, but little is known otherwise about this model.
The zero temperature zero-field phase diagram exhibits three distinct ground states, as illustrated in Fig.~\ref{fig:h0_phasediag},
and promises interesting physics also in finite fields, respectively at finite temperature. Here,
we shortly introduce these phases, with further details provided in the following sections.  In the case of large dimer
coupling $J_2$, the ground state is an exact 
product state formed by dimer singlet states on all the $J_2$ dimers,
while the remaining spins (referred to as monomer spins) are effectively decoupled. This leads to an extensive ground-state
entropy of $\ln(2)$ per unit cell in this regime ($J_2/J_1>2$), which is denoted the monomer-dimer (MD) phase. On the
other hand, for weak $J_2$, the system prefers to form 
dimer triplet states on all the $J_2$ dimers, while the
monomer spins predominantly orient themselves opposite to the polarization of the dimers.
This leads to a
ferrimagnetic state, akin to the  ferrimagnetic ground state of the mixed spin-1 and spin-1/2 model on the Lieb lattice~\footnote{This lattice appeared in the context of the CuO$_2$ planes of the high-$T_c$ superconductors \cite{Mattis88}
and thanks to extensions in Ref.~\cite{Lieb89} is now known as ``Lieb’' lattice.
Actually, an Ising version of this model had appeared in the literature already significantly earlier \cite{Fisher60a,Fisher60b},
but we will stick to established terminology.
}. Its
ferrimagnetic polarization follows from the Lieb-Mattis theorem~\cite{Lieb1962} in terms of the two different sublattices of the Lieb
lattice.
This phase is  therefore also denoted by ``LM'' in the following. These two phases, MD and LM,  are separated
by a further gapped phase, the dimer-tetramer (DT) phase, cf.\ Fig.~\ref{fig:h0_phasediag}. In this phase, two different kinds of local singlets form: besides the $J_2$-dimer singlets, also singlets on larger clusters with four spins are formed: namely, among the tetramers that are each composed of one $J_2$-dimer and its two neighboring $J_1$-coupled monomer spins. In the DT phase, the ground-state manifold is again highly degenerate and consists of all configurations of closed packings of tetramers, with the remaining $J_2$-dimers 
forming dimer (two-site) singlets.

We here examine the spin-1/2 Heisenberg antiferromagnet on the diamond-decorated square lattice in a magnetic field. In particular, we explore the ground-state phase diagram in the presence of a finite magnetic field as well as the thermal properties. 
For this purpose, we use a combination of analytical approaches and various computational methods, including exact diagonalization (ED), density matrix renormalization group (DMRG) calculations~\cite{White92,White93,Schollwoeck05}
and stochastic series expansion (SSE) QMC simulations~\cite{Sandvik1991,Sandvik1999,Syljuasen2002}, based on a dimer-decoupling of the Hamiltonian~\cite{Honecker16,Alet16}, in order to render the QMC sign-problem free.

After introducing the model in more detail in the following section \ref{sec:model}, we describe the
analytical and computational approaches that we used 
in Secs.~\ref{Analytical} and \ref{sec:methods}, respectively. Our results for the ground-state properties are presented in  Sec.~\ref{sec:GS}, and those at finite temperatures in Sec.~\ref{sec:finiteT}.
In passing, we provide reference data for the mixed spin-1/2 and spin-1 Heisenberg model
on the Lieb lattice, compare also App.~\ref{AppA}.
Finally, we provide our conclusions and future perspectives in Sec.~\ref{sec:conclusions}. 

\section{Model}\label{sec:model}
In the following, we consider the spin-1/2 Heisenberg
antiferromagnet on the diamond-decorated square lattice
in a magnetic field. The lattice is shown schematically in Fig.~\ref{fig:lattice} and  the Hamiltonian of the model is given by
\begin{eqnarray}
  H &=& J_1 \sum \limits_{i=1}^N \Bigl[ \mathbf{S}_{i,1} \cdot
  \Bigl(\mathbf{S}_{i,2}+\mathbf{S}_{i,3}+\mathbf{S}_{i,4}+\mathbf{S}_{i,5}
  \nonumber \\
  &&+
  \mathbf{S}_{i-\hat{x},2}+\mathbf{S}_{i-\hat{x},3}+\mathbf{S}_{i-\hat{y},4}+\mathbf{S}_{i-\hat{y},5}\Bigr)
  \Bigr]\nonumber \\
  &&+ J_2 \sum \limits_{i=1}^N \Bigl(\mathbf{S}_{i,2}\cdot \mathbf{S}_{i,3} +
  \mathbf{S}_{i,4} \cdot \mathbf{S}_{i,5}\Bigr)\nonumber \\
  &&- h\sum \limits_{i=1}^N\sum
  \limits_{\mu=1}^{5} S_{i,\mu}^z\,,
	\label{hamos}
\end{eqnarray}
where $\mathbf{S}_{i,\mu}= ({S}_{i,\mu}^x, {S}_{i,\mu}^y, {S}_{i,\mu}^z)$ represents the spin-1/2 operators assigned to the $\mu$-th spin within the $i$-th unit cell. We denote the corresponding lattice site  by $(i,\mu)$. Furthermore, the index $i-\hat x$ ($i-\hat y$) refers to the unit cell to the left (below) the $i$-th unit cell. 
Here, we consider a finite lattice with $N$ unit cells and $N_s=5N$ sites, imposing periodic boundary conditions, and with $N\rightarrow \infty$ in the thermodynamic limit (TDL). Typically, we use square lattices with $N=L^2$. Furthermore,  $J_1$ and $J_2$ are the two exchange interactions drawn in Fig.~\ref{fig:lattice} by black and red lines, respectively. The last term in $H$ accounts for the Zeeman coupling of the spin-1/2 particles to an external magnetic field $h$. 

The Hamiltonian (\ref{hamos}) can also be expressed in terms of the composite spins on the $2N$ dimers formed by the $J_2 $ bonds: In each unit cell $i$, a vertical dimer is formed by the spins  $\mathbf{S}_{i,2}$ and  $\mathbf{S}_{i,3}$, and the total dimer spin for this dimer $d$ is then 
$\mathbf{S}_{d}=\mathbf{S}_{i,2}+\mathbf{S}_{i,3}$. Similarly, the spins  $\mathbf{S}_{i,4}$ and  $\mathbf{S}_{i,5}$ form a horizontal dimer, and in this case
$\mathbf{S}_{d}=\mathbf{S}_{i,4}+\mathbf{S}_{i,5}$.
All total dimer spins represent locally conserved quantities with well defined quantum spin numbers. The remaining spins $\mathbf{S}_{i,1}$ are referred to as monomer spins. One can then express the Hamiltonian (\ref{hamos}) in a more compact form: 
\begin{eqnarray}
H\!&=&\!J_1\sum_{d=1}^{2N}\sum_{ (i,1) \in \mathcal{N}_d } \mathbf{S}_{d}\cdot\mathbf{S}_{i,1} + \frac{J_2}{2}\sum_{d=1}^{2N}\left(\mathbf{S}_{d}^2-\frac{3}{2}\right)\nonumber \\
&-&h\sum_{d=1}^{2N} S_{d}^z - h\sum_{i=1}^N S_{i,1}^z,
\label{hamDimer}
\end{eqnarray}
where  summations over $d$ extend over all the $2N$ dimers, and the inner sum of the first term extends over the two monomer
spins  $\mathbf{S}_{i,1}$ that are nearest neighbors of the $d$-th dimer (cf.\ Fig.~\ref{fig:lattice}), i.e., the lattice site $(i,1)$ is an element of the set of the two nearest-neighbor sites $\mathcal{N}_d$ of the $d$-th dimer. More specifically, for a vertical  (horizontal) dimer, these  are the two monomer spins to the left and right (top and bottom) of that dimer. 

The first term in the Hamiltonian (\ref{hamDimer})  corresponds to the  mixed spin-$S_d$ and spin-1/2 Heisenberg model on a Lieb lattice, whereas the second term provides a trivial shift of the energy depending on the quantum spin numbers $S_{d}$. Note that two different values of the quantum spin numbers $S_{d} = 0$, $1$ are available for the composite spin on each dimer, whereby the value $S_{d} = 0$ corresponds to a singlet-dimer state 
\begin{eqnarray}
|s\rangle_d=\frac{1}{\sqrt{2}}\left(|\!\uparrow\downarrow\rangle_{d}-|\!\downarrow\uparrow\rangle_{d}\right).
\label{dimer}
\end{eqnarray}
This leads to  a fragmentation of the effective mixed-spin Heisenberg models obtained from the Hamiltonian (\ref{hamDimer}) upon considering all possible combinations of quantum spin numbers $S_{d}$ for all the dimers. Hence, the ground state of the Heisenberg antiferromagnet on the diamond-decorated square lattice can be related to the lowest-energy eigenstates of the effective Heisenberg models (\ref{hamDimer})  taking into consideration  all available combinations of the quantum spin numbers $S_{d}$. In the following, we first introduce our methods and then explore the rich ground-state phase diagram of the Hamiltonian $H$,  shown further below in Fig.~\ref{fig:full_phasediag}.

\section{Exact analytical ground states}\label{Analytical}
We first consider the parameter regime with a dominant dimer coupling $J_2$, in which we can obtain exact analytical results for the ground state. More specifically,
for $J_2/J_1>2$ one can use the variational principle in order to derive an exact ground state of $H$ at zero field \cite{Morita2016}. The main idea of this approach consists in decomposing the  Hamiltonian into $4N$ cell Hamiltonians (this concrete decomposition is different from Ref.~\cite{Morita2016}):
\begin{eqnarray}
H=\sum_{d=1}^{2N}\sum_{(i,1)\in \mathcal{N}_d} H_{d,i},
\end{eqnarray}
with each cell Hamiltonian $ H_{d,i}$ corresponding to a single triangle involving one dimer $d$ and one of its two nearest-neighbor monomer spins, i.e., 
\begin{eqnarray}
H_{d,i} =\frac{J_2}{4}\left(\mathbf{S}_{d}^2-\frac{3}{2}\right)+J_1\mathbf{S}_{i,1}\cdot\mathbf{S}_{d} .
\label{hamTriangle}
\end{eqnarray}
Note that each dimer $d$ is part of   two triangles, leading to the additional factor of $1/2$ for the intra-dimer term proportional to $J_2$ in $H_{d,i}$ as compared to Eq.~\eqref{hamDimer}.

According to the variational principle~\cite{Shastry1981,Bose89,Bose90,Bose92}, the ground-state energy of $H$ has a {\it lower} bound, given by the sum of the lowest-energy eigenvalues 
$\varepsilon^{(0)}_{d,i}$ 
of the cell Hamiltonians (\ref{hamTriangle}),
\begin{eqnarray}
E_0&=&\langle\Psi_0|H|\Psi_0\rangle\nonumber\\
&=&\Bigl\langle \Psi_0\Bigr|\sum_{d=1}^{2N} 
\sum_{(i,1)\in \mathcal{N}_d} \!\!\! H_{d,i}\,\Bigl|\Psi_0\Bigr\rangle
\geq\sum_{d=1}^{2N}\sum_{(i,1)\in \mathcal{N}_d} \varepsilon_{d,i}^{(0)}\,. \
\end{eqnarray}
The energy-spectrum of each cell Hamiltonian $H_{d,i}$  can be expressed in terms of  quantum spin numbers $S_{t}$ and $S_{d}$ which are assigned to the composite spin operators $\mathbf{S}_{t}=\mathbf{S}_{d}+\mathbf{S}_{i,1}$ and $\mathbf{S}_{d}$, respectively, as follows,
\begin{eqnarray}
\varepsilon_{d,i}&=&-\frac{3}{8}(J_1+J_2)+\frac{J_1}{2}S_{t}(S_{t}+1) \nonumber \\
&&+\left(\frac{J_2}{4}-\frac{J_1}{2}\right)S_{d}(S_{d}+1)
\,.\quad
\end{eqnarray}
It is  straightforward to show  that for $h=0$ the eigenstate with quantum spin numbers $S_{t}=1/2$ and $S_{d}=0$ represents the true ground state of $H_{d,i}$ whenever $J_2/J_1>2$. Hence, 
in this regime  $\epsilon_{d,i}^{(0)}=-\frac{3}{8}J_2$. 
A finite field then simply leads to a polarization of the monomer spins, as long as it does not exceed a critical value.
Owing to this fact, the overall ground state of $H$ for $J_2/J_1>2$ and in the  monomer-dimer (MD) phase is 
\begin{eqnarray}
|\mathrm{MD}\rangle=\left\{ 
\begin{array}{l}
\prod_{i=1}^N|\sigma\rangle_{i,1}\otimes\prod_{d=1}^{2N}|s\rangle_d, \, \sigma\in\{\uparrow,\downarrow\},\, h=0 \ \\
\prod_{i=1}^N|\!\uparrow\rangle_{i,1}\otimes\prod_{d=1}^{2N}|s\rangle_d, \, h> 0
\end{array}\right. \
\label{monomerDimer}
\end{eqnarray}
which has the following energy 
\begin{eqnarray}
E_{\mathrm{MD}}/N=-\frac{3}{2}J_2-\frac{h}{2}.
\label{eMD}
\end{eqnarray} 
Note, that for $h=0$ the MD phase has an extensive ground-state degeneracy, $2^N$, as each of the $N$ monomer spins can be either in the up or down state. We will examine in Sec.~\ref{sec:GS} up to which field strength the MD phase is actually stable. 

The stability condition  $J_2/J_1>2$ of the MD phase at $h=0$ is in agreement with the results reported previously by Hirose \textit{et al}.~\cite{Hirose2017,Hirose2018,hirose}. They also verified the presence of the other exact ground state, referred to as the dimer-tetramer (DT) phase. 
The DT ground state of the spin-1/2 Heisenberg antiferromagnet on a diamond square lattice involves the singlet-dimer states $|s\rangle_d$ and  singlet-tetramer states $|t\rangle_d$, which are formed between a dimer $d$ and its two neighboring monomer spins, denoted $(i,1)$ and $(i',1)$ in the following:
\begin{eqnarray}
|t\rangle_d &=& \frac{1}{\sqrt{3}}(|\!\uparrow\rangle_{i,1}|\!\downarrow\uparrow\rangle_{d}|\!\downarrow\rangle_{i',1}+|\!\downarrow\rangle_{i,1}|\!\uparrow\downarrow\rangle_{d}|\!\uparrow\rangle_{i',1}) \nonumber \\
&&-\frac{1}{2}\left(|\!\uparrow\rangle_{i,1}|\!\uparrow\downarrow\rangle_{d}|\!\downarrow\rangle_{i',1} + |\!\uparrow\rangle_{i,1}|\!\downarrow\downarrow\rangle_{d}|\!\uparrow\rangle_{i',1}\right. \nonumber \\
&&+\left.|\!\downarrow\rangle_{i,1}|\!\uparrow\uparrow\rangle_{d}|\!\downarrow\rangle_{i',1}+ |\!\downarrow\rangle_{i,1}|\!\downarrow\uparrow\rangle_{d}|\!\uparrow\rangle_{i',1}\right)\, .
\label{ti}
\end{eqnarray}
In the DT phase, the highly degenerate ground-state manifold corresponds to the most dense packing of the singlet-tetramer states (\ref{ti}) on the diamond-decorated square lattice,
whereby one cannot accommodate more than ${N}/{2}$ singlet tetramers $|t\rangle_d$ on the diamond-decorated square lattice (the remaining dimers are in the singlet-dimer state $|s\rangle_d$). The ground-state energy in the DT phase is thus given by
\begin{eqnarray}
E_{\rm DT}/N = \frac{3}{2}\varepsilon_{s}+ \frac{1}{2}\varepsilon_{t}.
\label{edt}
\end{eqnarray}
Here, $\varepsilon_{s}=-\frac{3}{4}J_2$ refers to the energy of the singlet-dimer state $|s\rangle_d$, and $\varepsilon_{t}=-2J_1+\frac{J_2}{4}$ denotes the energy of the singlet-tetramer state $|t\rangle_d$. 
In order to obtain the actual stability regions of these two phases for finite fields, we turn to computational methods. 

\section{Computational Approaches}\label{sec:methods}
For our further analysis of the phase diagram of the spin-1/2 Heisenberg diamond-decorated square lattice as well as its thermodynamic properties, we have used a combination of various computational approaches. In this section, we provide some details regarding the application of these different methods to the model considered here. 

\subsection{DMRG}
\label{sec:DMRG}
%
The ED and QMC simulations to be presented in the next subsections indicate that there is one important class of ground states that are not captured by the MD and DT wave functions discussed in the previous section: the particular choice $S_d = 1$ for all $2N$ dimers. This amounts to an effective mixed spin-1 and spin-1/2 Heisenberg model on a Lieb lattice, given by the Hamiltonian (\ref{hamDimer}).
For $h=0$, the effective Hamiltonian reads
\begin{eqnarray}
H^\mathrm{LM}_\mathrm{eff} = J_1\sum_{d=1}^{2N}\sum_{(i,1)\in \mathcal{N}_d} \mathbf{S}_{d}\cdot\mathbf{S}_{i,1} + \frac{J_2}{2}N.
\label{hamLieb}
\end{eqnarray}
In contrast to the  case of fixed dimer-singlet states, the Hamiltonian (\ref{hamLieb}) cannot be solved analytically and we have therefore adopted the DMRG method implemented in the Algorithms and Libraries for Physics Simulations (ALPS) project \cite{baue11} in order to find its lowest-energy eigenstates. For this purpose, we have performed DMRG calculations taking into account up to 2000 kept states and  up to 20 sweeps for lattices with up to $N=36$ unit cells with periodic boundary conditions. The respective lowest-energy eigenvalue of the spin-1/2 Heisenberg antiferromagnet on a diamond-decorated square lattice is given for $h=0$ by the equation:
\begin{eqnarray}
E_\mathrm{LM}=E_{\rm L}+ \frac{J_2}{2}N,
\label{elm}
\end{eqnarray}
where $E_{\rm L}$ denotes the ground-state energy 
of the mixed spin-1 and spin-1/2 Heisenberg model on 
the corresponding Lieb lattice with $N$ unit cells at zero magnetic field.  
According to the Lieb-Mattis theorem~\cite{Lieb1962}, the lowest-energy eigenstate of the mixed spin-1 and spin-1/2 Heisenberg model on a Lieb lattice in a zero field belongs to the sector with the total spin given by the absolute value of the difference of the total spin on the two sublattices $S=|S_A-S_B|$. For the Lieb lattice composed of $N=36$ unit cells we have indeed obtained a ferrimagnetic ground state with  total spin $S=|S_A-S_B|=|72-18|=54$ and  energy $E_{L}=-88.5600047 J_1$, i.e., the ground-state energy $\varepsilon_\mathrm{L}=-2.46000J_1$ per unit cell. We note that this value compares well to the value $\varepsilon_\mathrm{L}=-2.46083 J_1$ for the ground-state energy of the mixed spin-1 and spin-1/2 Heisenberg antiferromagnet on the Lieb lattice in the TDL, given in Ref.~\cite{Hirose2018}.

In order to construct the ground-state phase diagram we have compared the  energies (\ref{eMD}),(\ref{edt}) and (\ref{elm}), which were obtained either by analytical or by numerical calculations of a lattice with $N=36$ unit cells. 
To study the magnetization process and thermodynamic quantities in finite magnetic field, all energies in zero field are shifted by the Zeeman term according to the formula 
\begin{eqnarray}
E(m_\mathrm{tot}, N, h) = E(m_\mathrm{tot}, N, h=0)-h\: m_\mathrm{tot} \, , \
\label{einnonzero}
\end{eqnarray}
where $m_\mathrm{tot}$ are the eigenvalues of $S_\mathrm{tot}^z=\sum_{i=1}^N\sum_{\mu=1}^5 S^z_{i,\mu}$. 
Field-driven changes of the lowest-energy eigenstates from the  sectors with the total spins 
$m_\mathrm{tot}$ and $m_\mathrm{tot}'$
are obtained from~\cite{hone00}
\begin{eqnarray}
h = \frac{E(m_\mathrm{tot}, N, h=0)-E(m_\mathrm{tot}', N, h=0)}{m_\mathrm{tot}-m_\mathrm{tot}'}.
\label{pole}
\end{eqnarray}
We note that within the LM phase, the ground-state energy in the TDL and finite $h$
is given by
\begin{equation}\label{eLM}
    E_\textrm{LM}/N=\varepsilon_\mathrm{L}  +\frac{J_2}{2}-\frac{3}{2}h,
\end{equation}
a result that will be useful further below. 

\subsection{Exact Diagonalization}
In addition to DMRG, we also performed exact diagonalizations of the Hamiltonian $H$ on systems with up to $N_s=30$ spins.
First, we exploit conservation of the local spin of each dimer by expressing the Hamiltonian in the form (\ref{hamDimer}). Thus, the problem boils down to diagonalizing the Hamiltonian for a given configuration of total dimer spins $S_{d} = 0$, $1$ \cite{DRHS07,Honecker_2011,PhysRevB.86.054412,Honecker16}.
In fact, it suffices to perform this computation for each topologically inequivalent pattern. We have identified the inequivalent patterns by computer enumeration. For example, for $N_s=30$ spins (6 unit cells), we find 178 inequivalent configurations of the dimer spins $S_{d}$. Then, we need to take the degeneracy of the corresponding configurations into account. For example, the configurations with all $S_{d} = 0$ or
all $S_{d} = 1$ are unique, and generally there are $2N_s/5$ configurations with exactly one $S_{d_i} = 1$. Furthermore, in the case of the $N_s=30$ system, there are up to 120 different realizations of a given pattern with an intermediate number of dimer triplets.

To diagonalize each of these cases, we first use conservation of  $S^z_\mathrm{tot}$, as well as spin inversion. We further use $SU(2)$ symmetry to reconstruct the sector with $m_\mathrm{tot}=1$ from the other ones. Thermodynamic quantities such as the specific heat and magnetic susceptibility can then be computed from the eigenvalues and the associated quantum numbers. In  sectors where we have so many $S_{d} = 1$ dimers that the Hilbert space dimensions become large, we also use the remaining spatial symmetries of the configuration to further block-diagonalize the system.
The largest matrix to be diagonalized then occurs in the sector with $m_\mathrm{tot}=2$ and for all dimers in the triplet configuration; for $N_s=30$ the resulting maximal dimension is $257304 \approx 2.6\cdot 10^5$, which is considerably smaller than the total dimension $2^{30} \approx 10^9$ of this system. Still, this significantly exceeds the size of a previous computation~\cite{PhysRevB.74.020403} where we had used
a custom diagonalization routine~\cite{honecker2008openmp}, while the present diagonalization is instead carried out with a recent version of the Intel$^\circledR$ Math Kernel Library.

At the end of this procedure, the full spectrum can be reconstructed for any value of $J_2$, $J_1$ and $h$ thanks to the conservation of the total spin on the dimers and $z$-component of the total spin $m_\mathrm{tot}$. Thus, we can evaluate thermodynamic properties for all $(J_2/J_1, h/J_1, T/J_2)$ by post-processing the results of a single diagonalization run for a given system size $N_s$.

\subsection{Quantum Monte Carlo}
In order to study the thermodynamic properties of the spin-1/2 Heisenberg antiferromagnet on the diamond-decorated square lattice on system sizes that extend beyond those accessible to exact diagonalization, we make use of QMC simulations. In the following we comment on the QMC method that  we used for this purpose. 

The SSE QMC method with directed loops \cite{Sandvik1991,Sandvik1999,Syljuasen2002} offers a highly efficient and unbiased approach to study quantum spin models. However, 
introducing geometric frustration while working in the conventional local spin-$S^z$ basis generally leads to a
sign problem, i.e., an exponential drop of the statistical accuracy of the QMC simulations at low temperatures and large system sizes~\cite{Troyer2005,Hangleiter2020,Hen2021}. 
Fortunately, in certain cases, this issue can be eliminated when performing the QMC simulations in a basis different from the local spin-$S^z$ basis. More specifically, one considers instead appropriate basis states after decomposing the Hamiltonian into separate terms of few-sites clusters, such as dimers or trimers~\cite{Honecker16,Alet16,Weber2021}. 
The case of dimers can be used to eliminate the sign problem completely for, e.g.,  the fully frustrated bilayer model~\cite{Alet16,NgYang17,Stapmanns2018}, while a local spin-trimer basis avoids the sign-problem for the fully frustrated trilayer \cite{Weber2021} $S=1/2$ antiferromagnet. For the diamond-decorated square lattice  considered here, a finite value of the  coupling $J_2$ leads to geometric frustration. We can avoid the associated sign problem that persists when using the local spin-$S^z$ basis, by treating all $J_2$-dimer spins in the spin-dimer basis, while leaving the local $S^z$-basis to the monomer spins. In this combined 5-site basis for each unit cell, the Hamiltonian $H$ can be simulated free of a sign problem, using the SSE approach based on the  abstract operator loop update introduced in Ref.~\cite{Weber2021}. During the operator-loop update of the SSE simulations, binary operators (such as the bit-wise exclusive-or operation) are used in a binary representation of the local cluster states. We refer to Refs.~\cite{Weber2021,Weber2022} for further details on this QMC approach. Here, we performed QMC simulations for systems with values $L$ up to $24$. 


\section{Ground-state phase diagram}\label{sec:GS}
In the following we describe in detail the  ground-state phase diagram of the spin-1/2 Heisenberg antiferromagnet on the diamond-decorated square lattice, described by the Hamiltonian $H$, up to high magnetic fields, as obtained from our combination of DMRG, exact diagonalization as well as analytical results. 

\begin{figure}[t]
    \centering
    \includegraphics[width=0.99\columnwidth]{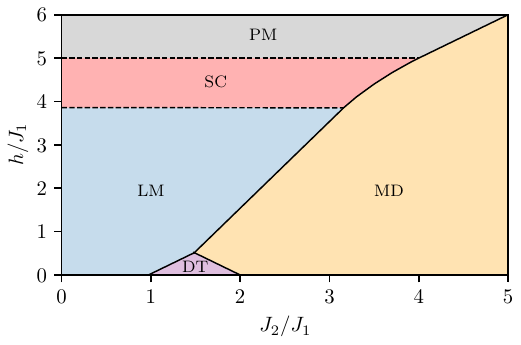}
    \caption{Ground-state phase diagram of the spin-1/2 Heisenberg antiferromagnet on the diamond-decorated square lattice in the $J_2/J_1-h/J_1$ plane, containing the Lieb-Mattis (LM), dimer-tetramer (DT), monomer-dimer (MD), spin-canted (SC) and the saturated paramagnetic (PM) phase. Dashed (solid) lines denote continuous (discontinuous) quantum phase transitions.}
    \label{fig:full_phasediag}
\end{figure}

\begin{figure}
    \centering
    \includegraphics[width=0.99\columnwidth]{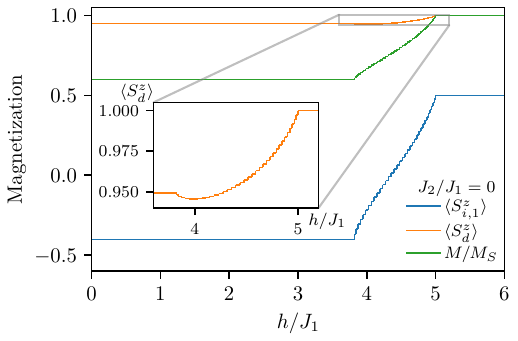}
    \caption{Local dimer magnetization $\langle S_d^z \rangle$, local monomer magnetization $\langle S_{i,1}^z \rangle$ and total magnetization $M$ (divided by the saturated magnetization $M_S$) at zero temperature, as functions of magnetic field $h/J_1$ for $J_2=0$ as obtained from DMRG for a $L=6$ system.}
    \label{fig:scphase}
\end{figure}

First, we review the phases that appear at zero field $h=0$. The ground-state phase diagram in this limit has been obtained
in previous works~\cite{Hirose2016,Hirose2017,Hirose2018} and it features three distinct phases -- LM, DT and MD, depending
on the coupling ratio $J_2/J_1$, cf.\ Fig.~\ref{fig:h0_phasediag}. In the LM regime, the ground-state energy is minimized by all $J_2$-dimers being in the 
triplet state, while the monomer spins are oriented predominantly  opposite to the dimer spins. Note that the quantum nature of the LM ferrimagnetic state is well observable by the reduction of the local magnetizations of the monomer and dimer spins as shown in Fig.~\ref{fig:scphase}. While  quantum fluctuations reduce the magnetization of the monomer spins by  approximately 20\%,  the quantum reduction of the magnetization of the dimer spins is much more subtle -- only about of 5\%. In agreement with the LM theorem, both local magnetizations are consistent with the total magnetization per site of $M/M_S=3/5$ (where $M_S$ denotes full saturation) that is inherent to the LM ferrimagnetic phase. For $J_2/J_1$ between about 0.974  and 2, 
the zero-field ground state is the highly degenerate DT phase, characterized by a dense packing of the singlet-tetramer states $|t\rangle_d$ given by Eq.~(\ref{ti}), while the remaining $J_2$-dimers reside in a singlet-dimer state (\ref{dimer}). In the highly frustrated parameter region $J_2/J_1>2$ the MD phase (\ref{monomerDimer}) is realized in the ground state, which was described in detail in Sec.~\ref{Analytical}. 

The ground-state phase diagram including the  magnetic field $h$ is shown in Fig.~\ref{fig:full_phasediag} in the $J_2/J_1-h/J_1$ 
plane. As one can see from Fig.~\ref{fig:full_phasediag}, the LM phase is stable  up to about $h/J_1\approx 4$ and  spreads
out to larger  interaction ratios $J_2/J_1$ with increasing  magnetic field. By contrast, the MD phase (\ref{monomerDimer}) 
extends towards lower interaction ratio $J_2/J_1<2$ in finite magnetic fields as compared to the parameter regime that is
accessible to the variational approach, cf.\ Sec.~\ref{Analytical}. We also find that for $J_2/J_1\geq 4$, the MD phase is stable all the way up to the saturation field $h_\mathrm{sat}=J_1+J_2$, beyond which the fully polarized, saturated paramagnetic (PM) regime is entered. 
On the other hand, the DT phase narrows quickly for finite magnetic fields and it disappears completely at
$h/J_1 \approx 0.5$. 

In addition to  the LM, DT and MD phases, the phase diagram in Fig.~\ref{fig:full_phasediag} exhibits  two high-field phases. Besides the fully polarized, saturated PM regime, we identify a spin-canted (SC) phase, with a continuously rising magnetization upon increasing the magnetic field.
As shown in Fig.~\ref{fig:scphase}, inside the SC phase the local monomer spins continuously align with the magnetic field upon increasing the field strength. Initially, the local dimer magnetization decreases slightly, before it eventually increases to  full polarization  as well. Qualitatively, this behavior is well captured by the classical Heisenberg model of the mixed spin-1 and spin-1/2 model Heisenberg model on the underlying Lieb lattice, as detailed in  App.~\ref{AppA}: Within the SC phase, the spins are canted with respect to the magnetic field direction, displaying biconical structures, resembling those found in, e.g., the classical anisotropic Heisenberg model at finite magnetic fields~\cite{Fisher75, Holtschneider2007}. Fig.~\ref{fig:full_phasediag} furthermore shows that  the SC phase is separated from the PM and the LM  phase by continuous field-driven quantum phase transitions. In contrast,  all other field-driven phase transitions between the various ground-state phases are discontinuous.

\begin{figure}
    \centering
    \includegraphics[width=0.99\columnwidth]{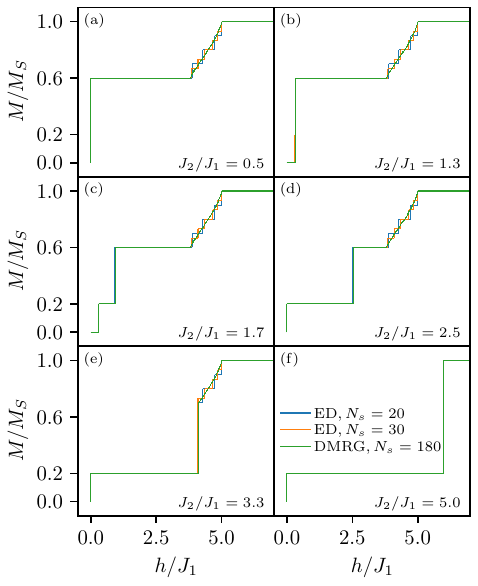}
    \caption{Zero-temperature magnetization curves of the spin-1/2 Heisenberg antiferromagnet on the diamond-decorated square lattice as obtained from full exact diagonalization and DMRG method for system sizes $N_s=20$, $N_s=30$ and $N_s=180$, respectively, for several values of the interaction ratio $J_2/J_1$. 
    In all plots, the total magnetization $M$ is scaled with respect to its saturated value $M_S$.}
    \label{fig:dmrg}
\end{figure}

The phase diagram and the nature of the transitions in Fig.~\ref{fig:full_phasediag} can be directly identified from the  
zero-temperature magnetization curves for the Hamiltonian $H$. These are shown in Fig.~\ref{fig:dmrg} along several vertical cuts through the ground-state phase diagram. The magnetization curves presented in Fig.~\ref{fig:dmrg} were obtained by two different numerical methods: by exact diagonalization for $N_s=20, 30$ and by the DMRG method supplemented with exact analytical results for $N_s=180$, respectively. Overall, the results obtained from both methods are in excellent agreement, taking into account that within the SC phase, the smaller size of the system either with $N_s=20$ or $N_s=30$ leads to more pronounced discrete steps in the stair-case profile of the magnetization. 

More specifically, for an interaction ratio of $J_2/J_1=0.5$, the magnetization exhibits an extended 3/5-plateau at
low fields, characteristic of the ferrimagnetic LM phase, followed by a stair-case increase of the  magnetization, which 
evolves into a continuous magnetization increase in the TDL, and  eventually terminates in the fully saturated PM phase at a
magnetic field of $h/J_1=5$ [cf.\ \ref{fig:dmrg}(a)]. Note that the jump of $M$ upon approaching the zero-field limit is not a numerical artifact but reflects the ferrimagnetic nature of the LM phase, i.e., the immediate response to an infinitesimal field.

Similarly, in agreement with the ground-state phase diagram, there exists a zero-magnetization plateau for $J_2/J_1=1.3$ and $J_2/J_1=1.7$ in Figs.~\ref{fig:dmrg}(b) and \ref{fig:dmrg}(c) inherent to the gapped DT state (closer inspection resolves a tiny $1/5$-plateau on the $N_s=30$ system in Figs.~\ref{fig:dmrg}(b) -- a
finite-size effect on this particular cluster). 
For larger values of  $J_2/J_1>2$,  shown in Figs. \ref{fig:dmrg}(d), \ref{fig:dmrg}(e) and \ref{fig:dmrg}(f), the zero-magnetization plateau disappears. Instead, here, the monomer spins become fully polarized within the MD phase already for an arbitrarily weak, finite magnetic field. This results in the jump of $M$  upon approaching the zero-field limit and the immediate onset of the intermediate 1/5-plateau, characteristic of the MD phase. 
While for the lower value of $J_2/J_1=2.5$ the 3/5-plateau of the LM phase, and a subsequent steady increase  of the magnetization in the SC phase can be observed, the magnetization curve for the higher value of $J_2/J_1=3.3$ exhibits a discontinuous field-driven transition from the MD phase (1/5-plateau) into the SC phase. Finally, the magnetization curve for sufficiently high values of $J_2/J_1 \geq 4$ shows a direct  jump of the magnetization from the 1/5-plateau of the MD phase towards the fully saturated PM regime.      

\section{Thermal Properties}\label{sec:finiteT}
In the following, we will investigate several aspects of the  thermal properties of the spin-1/2 Heisenberg model on the diamond-decorated square lattice in the presence of a magnetic field, 
focusing on the  different regions of the ground-state phase diagram, which were  detailed in the previous section.
\subsection{Thermodynamics in the MD regime}
\begin{figure}
    \centering
    \includegraphics[width=0.99\columnwidth]{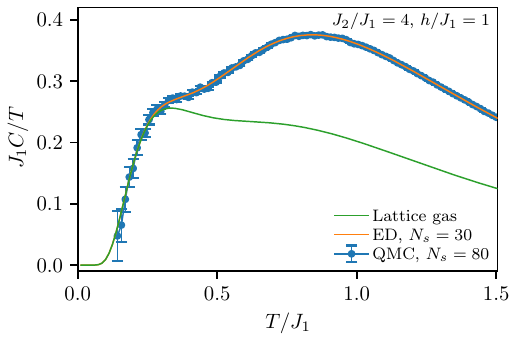}
    \caption{Specific heat $C$ divided by the temperature $T/J_1$ as a function of temperature in the MD phase at an interaction ratio of $J_2/J_1=4$ and a magnetic field of $h/J_1=1$ as obtained from full exact diagonalization, QMC as well as the effective lattice-gas model.}
    \label{fig:MD_TD}
\end{figure}

\begin{figure}[t]
    \centering
    \includegraphics[width=0.99\columnwidth]{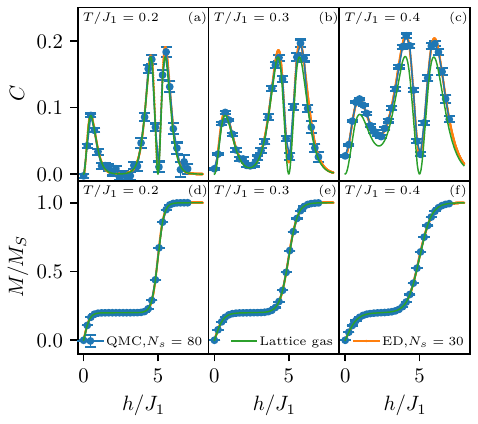}
    \caption{Thermodynamic properties in the MD phase at interaction ratio $J_2/J_1=4$ as a function of magnetic field $h/J_1$ obtained from full exact diagonalization, QMC as well as the effective lattice-gas model. Top row: Specific heat $C$ for temperatures: (a) $T/J_1=0.2$, (b) $T/J_1=0.3$, (c) $T/J_1=0.4$. Bottom row: Magnetization $M$, divided by the saturation magnetization $M_S$ for temperatures: (d) $T/J_1=0.2$, (e) $T/J_1=0.3$, (f) $T/J_1=0.4$  }
    \label{fig:LG_comp}
\end{figure}

We start  by investigating the thermodynamic properties in the MD regime.
Here, a simple lattice-gas model~\cite{DRHS07,Strecka17,Strecka22} can be used to describe the 
relevant low-energy excitations 
in the regime $J_2/J_1 \geq 4$. 
In addition to a set of  free $S=1/2$ spins, corresponding to the monomer spins $S_{i,1}$ in a magnetic field, this model contains a lattice gas of hard-core particles that correspond to the dimer-singlet states on the $J_2$-dimers. These particles describe localized magnons, i.e., a flat band of magnetic excitations, relative to the fully polarized state~\cite{Schulenburg02,DRHS07,Strecka17,Strecka22}. 
The lattice-gas model is given, up to a constant,  by the effective Hamiltonian
\begin{equation}
  H^\mathrm{MD}_\mathrm{eff} = - h\sum \limits_{i=1}^N S_{i}^z - \mu \sum\limits_{i=1}^{2N} n_{d}  \, ,
	\label{hamef}
\end{equation}
where $n_{d}\in{0,1}$  denotes the local occupation number of the hard-core particles. A value of $n_d=1$ ($n_d=0$)  corresponds to the presence of a singlet (triplet) state on the $J_2$-dimer $d$.
The chemical potential $\mu=J_1+J_2-h$ is given by the energy difference between the singlet and lowest-energy triplet state
on the lattice.
All thermodynamic properties then follow from the free energy
\begin{eqnarray}
  \frac{F}{N} &=& 2J_1 +  \frac{J_2}{2} - 2h
   - T \ln\left[2 \cosh \left( \frac{h}{2T} \right) \right]
\nonumber \\
&& - 2 T \ln\left[ 1 + \exp\left(\frac{\mu}{T}\right) \right] \, ,
\label{eq:MDthermo}
\end{eqnarray}
where the constant ensures that, at $T=0$, the ground-state energy detailed in Sec.~\ref{Analytical} is recovered,
including Eq.~(\ref{eMD}) for $\mu > 0$.
The ground state corresponds to a fully occupied (empty) lattice of singlets below (above) the saturation field $h_\mathrm{sat}=J_1+J_2$. One can  approximately describe the thermodynamic properties in the MD regime also by a spin model that accounts for all the  dimer states (this model corresponds to the limit $J_1=0$ of $H$). However, the above lattice-gas model already turns out to describe the low-temperature thermodynamics remarkably well.   
Indeed, Figs.~\ref{fig:MD_TD} and \ref{fig:LG_comp} show that the lattice-gas model describes the thermodynamic properties at low temperatures rather accurately, up to $T/J_1\approx 0.4$, where the specific heat starts to show noticeable deviations to the data obtained using exact diagonalization ($N_s=30$) and QMC ($N_s=80$), indicating that at higher temperatures additional excitations become relevant (note that over the whole regime there is an excellent agreement between the $N_s=30$ ED data and the $N_s=80$ data obtained from QMC).
From the phase diagram in Fig.~\ref{fig:full_phasediag} we expect additional excitations to become most relevant in the regime near $h/J_1\approx 5$ at $J_2/J_1=4$, due to the SC phase. Indeed, in Fig.~\ref{fig:LG_comp}(c), for $T/J_1=0.3$, small differences are already resolved in this magnetic-field range.

The magnetization is less susceptible to these additional states, and exhibits an excellent agreement with the numerical data for all temperatures considered here. It is noteworthy that in all cases, the results obtained from  exact diagonalization and QMC agree very well with each other, indicating that the system sizes considered for the exact diagonalization are already representative of  the TDL in this regime. Our analysis therefore indicates that in addition to the exact analytical results for the MD ground state, also the low-temperature thermal properties can be understood analytically by means of a simple effective lattice-gas model given by the Hamiltonian (\ref{hamef}).

\subsection{Thermodynamics in the LM and SC regime}
\begin{figure}[t]
    \centering
    \includegraphics[width=0.99\columnwidth]{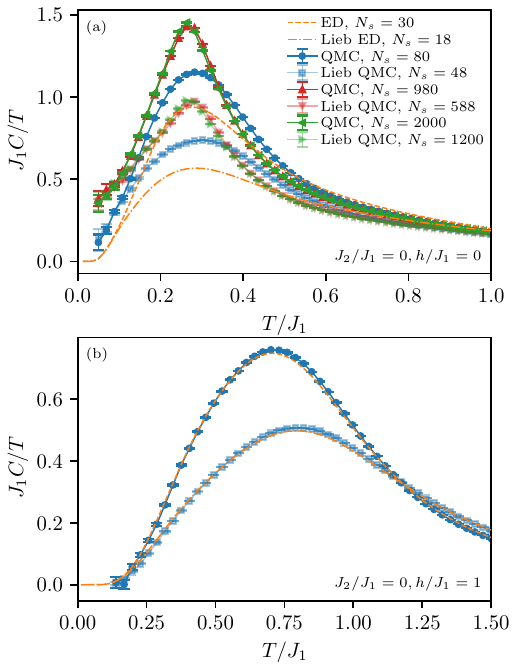}
    \caption{Specific heat $C$ divided by temperature $T/J_1$ as a function of
    temperature at magnetic field $h/J_1=0$ (top panel) and $h/J_1=1$ (bottom
    panel). We compare results for the diamond-decorated square lattice model for $J_2=0$,  with results on the mixed-spin Lieb lattice. For comparison, the Lieb lattice data are scaled by a factor of $3/5$ to account for the larger unit cell on the diamond-decorated square lattice.}
    \label{fig:LM_TD}
\end{figure}

Next, we consider the thermodynamic properties in the LM and SC phase. In the LM phase, we expect the low-temperature thermodynamics to be governed by the underlying mixed-spin Lieb lattice. For a quantitative comparison we considered the case of 
zero interaction $J_2=0$. In Fig.~\ref{fig:LM_TD}, we show the specific heat $C$ for the diamond-decorated square lattice model as obtained from ED (for $N_s=30$ sites) and QMC
as well as for the corresponding mixed-spin Lieb lattice model.
We observe that in both cases, $h=0$ and $h=J_1$, the mixed-spin Lieb lattice model captures the low-temperature asymptotic behavior, while the behavior differs noticeably at intermediate temperatures. 
This deviation is due to additional contributions with singlet configurations for the diamond-decorated square lattice.
These additional states are located at higher energies, but they have a high density such that they lead to a relevant
contribution to $C$ in the temperature window of Fig.~\ref{fig:LM_TD}.

Another point concerns the strong finite-size effects in Fig.~\ref{fig:LM_TD}(a). These are due to the $h=0$ ground state being a
spin-$3 N_s/10$ multiplet (compare Sec.~\ref{sec:DMRG}) such that on a finite-size system, part of the entropy is located at $T=0$.
The case $N_s=2000$ ($1200$) should be a good approximation to the TDL,
as is indicated by comparison with the $N_s=980$ ($588$) data.
Indeed, these data for $C/T$ in Fig.~\ref{fig:LM_TD}(a)
approach a constant for $T\to0$, as expected for a ferro- or ferrimagnet in two dimensions, while
the activated low-temperature asymptotics of the $N_s=30$ ED data reflects a finite-size gap of about $0.434 J_1$.

Applying a magnetic field $h=J_1$ opens a gap of the same size in the excitation spectrum. This leads to activated low-temperature behavior and negligible finite-size, as the good agreement of the ED and QMC results for $N_s = 30$ and $80$ ($18$ and $48$) in Fig.~\ref{fig:LM_TD}(b) shows.
As in the case $h=0$, we again observe significant differences between the diamond-decorated square lattice model and the mixed-spin Lieb lattice throughout most of Fig.~\ref{fig:LM_TD}(b) with the
exception of the region $T \lesssim 0.2 J_1$, where we observe the exponentially activated low-temperature asymptotics. This deviation can again be attributed to the large number of additional contributions with singlet configurations for the diamond-decorated square lattice that we already mentioned in the context of Fig.~\ref{fig:LM_TD}(a).

\begin{figure}[t]
    \centering
    \includegraphics[width=0.99\columnwidth]{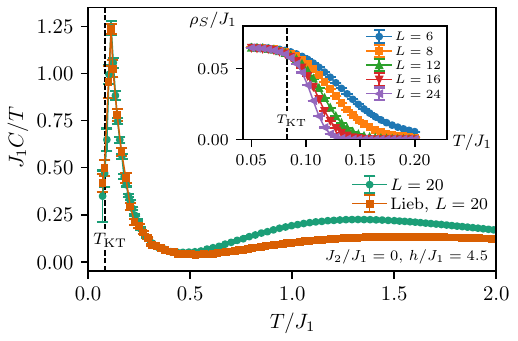}
    \caption{Specific heat $C$ divided by the temperature $T/J_1$ as a function of temperature in the SC phase for $J_2/J_1=0$ and $h/J_1=4.5$ as obtained from QMC. The dashed vertical line denotes the KT transition temperature $T_\mathrm{KT}$, given in the main text. The corresponding data for the mixed-spin Lieb lattice model is shown for comparison as well, and has been rescaled by a factor of $3/5$ to account for the larger unit cell on the diamond-decorated square lattice. 
    The inset shows the temperature dependence of the spin stiffness of the diamond-decorated square lattice model for the same parameters for different system sizes, along with the dashed line indicating $T_\mathrm{KT}$.}
    \label{fig:SC_TD}
\end{figure}

Figure \ref{fig:SC_TD} presents similar results for $C/T$ at $h/J_1 = 4.5$,
i.e., a point in the SC phase.
There is a maximum at temperatures $T > J_1$ that, like for Fig.~\ref{fig:LM_TD}(b), is affected by only small finite-size effects. Furthermore, also like in Fig.~\ref{fig:LM_TD}, the
diamond-decorated square lattice model and the mixed-spin Lieb lattice differ in this high-temperature
region. There is a second low-temperature maximum for $T$ just above $0.1 J_1$ that is remarkably well
captured by the mixed-spin Lieb lattice. This low-temperature maximum in $C/T$ is affected
by stronger finite-size effects,
but we checked that the residual finite-size effects on the $L=20$ data shown in Fig.~\ref{fig:SC_TD} are not substantial. 

We 
recall that the SC phase is characterized by (quasi-)long-range antiferromagnetic (XY) order in the plane perpendicular to
the magnetic-field direction at zero (low) temperatures, as discussed in the previous section. At non-zero temperatures, this leads to
a (Berezinskii-)Kosterlitz-Thouless (KT) transition~\cite{Berezinskii1971,Kosterlitz1973,Kosterlitz1974} at a finite temperature $T_\mathrm{KT}$. Beyond $T_\mathrm{KT}$, the XY quasi-long range order is destroyed by the proliferation of vortex excitations.
We have estimated $T_\mathrm{KT}$ using a standard finite-size scaling analysis of the spin stiffness $\rho_S$, as obtained from the spin winding number fluctuations~\cite{Weber1988,Harada1998} (cf.\ App.~\ref{AppB} for details). The inset of Fig.~\ref{fig:SC_TD} shows the temperature dependence of $\rho_S$ for $J_2/J_1=0$ and $h/J_1=4.5$ for increasing system sizes, exhibiting a drop near 
$T/J_1\approx 0.08$. 
From a quantitative analysis~\cite{Weber1988,Harada1998} of the QMC data, along the line $h/J_1=4.5$, the  KT transition temperature is found to be $T_\mathrm{KT}/J_1\approx 0.0825$ across the SC regime (cf.\ App.~\ref{AppB} for details). The specific heat $C$ displays a maximum at a temperature slightly above the KT transition temperature, as typical for the KT transition, associated to the entropy release from  vortex unbinding \cite{ChaikinLubensky1995}. This is  shown explicitly in Fig.~\ref{fig:SC_TD} for the interaction $J_2=0$ and the magnetic field $h/J_1=4.5$.

\subsection{Thermal LM-MD phase boundary}
Finally, we turn to consider the thermal properties within the parameter regime where the transition between the LM and MD phases takes place 
at zero temperature. As detailed in Sec.~\ref{sec:GS}, in the presence of a finite magnetic field $h \gtrsim 0.5J_1$, the LM and MD states at $T=0$ are separated by a direct discontinuous quantum phase transition line. Across this line, 
the $J_2$-dimer states change from triplets in the LM phase to singlet states in the MD phase (the monomer spins are fully polarized along the magnetic field in the MD phase, while their mean value is reduced due to quantum fluctuations in the LM phase).

Recently, such discontinuous quantum phase transitions in coupled spin-dimer and spin-trimer systems were examined in other models~\cite{Stapmanns2018,Jimenez2021,Weber2021}, and it was shown that first-order thermal phase transitions emerge out from  the discontinuous quantum phase transition line, terminating in a line of thermal critical points. Moreover, these thermal critical points belong to the  two-dimensional Ising universality class, in accord with the binary variable associated to the presence/absence of a singlet state on the spin dimers (such as the variable $n_d$ introduced above). Here, the LM-MD transition line offers another realization for such a scenario.  We thus examine it in more detail. 

\begin{figure}[t]
    \centering
    \includegraphics[width=0.99\columnwidth]{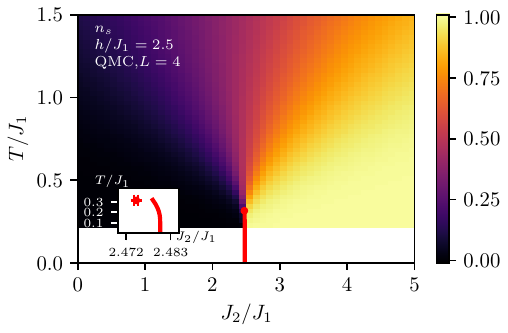}
    \caption{Mean singlet occupation in the vicinity of the LM-MD transition for $h/J_1=2.5$ as obtained from QMC for the  system size $L=4$. The red line shows the first-order transition line  obtained from comparing the free energies of both phases, extended up to the location of the critical point (symbol), as extracted from a finite-size analysis of the QMC data (see text for details). In the low-$T$ region (white), the QMC data exhibit large statistical fluctuations and have been cut off. }
    \label{fig:h_25_nd}
\end{figure}

As an example, Fig.~\ref{fig:h_25_nd} shows the mean singlet occupation 
\begin{equation}
n_s=\left\langle \frac{1}{2N}\sum_{d=1}^{2N} n_d \right\rangle
\end{equation}
of the $J_2$-dimers as a function of $J_2$ along a cut at constant $h/J_1=2.5$ across the LM-MD transition region. At low temperatures, this quantity exhibits a jump from a  value of 0 to a value of 1 upon increasing $J_2/J_1$ across the quantum phase transition near $J_2/J_1\approx 2.5$. For temperatures beyond about $0.3J_1$, we instead observe a smooth variation of $n_s$ with increasing $J_2$. This already provides indication for the existence of a low-$T$ discontinuous thermal phase transition line and its termination in a critical point.  While the precise position of the critical point needs to be extracted from QMC simulations (as detailed below), 
the first-order transition line at finite temperature can be estimated by simply comparing the free energies of both phases, following the approach used in Refs.~\cite{Weber2021,Stapmanns2018}. 

 Both the LM and MD phase have a finite excitation gap atop their respective ground states. 
Here, 
we  therefore use a generic estimate for the free energy of a gapped system at low $T$
that in the relevant parameter regime is given by
\begin{equation}
    \frac{F}{N} =-\frac{1}{N} \: T \ln Z  \approx  \frac{E_0}{N} - 2 T \ln (1 + {\rm e}^{-\Delta/T} ) \, , \label{eq:F_gap}
\end{equation}
where $E_0$ is the ground-state energy and $\Delta$ the excitation gap.
The factor two in front of the logarithm in Eq.~(\ref{eq:F_gap}) accounts for the two dimers
in the unit cell (note that triplets get polarized in a magnetic field and thus no spin-degeneracy factors enter Eq.~(\ref{eq:F_gap})).
We note that in the MD phase and for $h > J_1 + J_2$, Eq.~(\ref{eq:F_gap}) amounts to a low-temperature approximation
of the exact expression (\ref{eq:MDthermo}) for the effective lattice-gas model
with $\Delta = -\mu$.
At a fixed magnetic field, the transition line is then obtained from the points $J_2^c(T_c)$, for which the coexistence condition $F_\mathrm{LM}(J_2^c,T^c) = F_\mathrm{MD}(J_2^c,T^c)$ holds. Based on Eq.~(\ref{eq:F_gap}), we expect the first-order line not to be vertical, but to bend towards the phase with the larger excitation gap.

\begin{figure}[t]
    \centering
    \includegraphics[width=0.99\columnwidth]{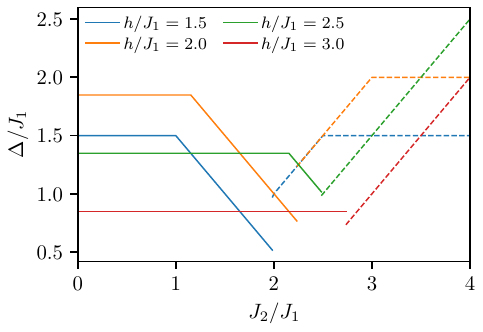}
    \caption{Excitation gap $\Delta/J_1$ as function of the interaction ratio $J_2/J_1$ for various magnetic fields $h/J_1$ as obtained from  exact diagonalization for the $N_s=30$ system. Solid (dashed) lines are used to denote the excitation gap in the LM (MD) regime.}
    \label{fig:gaps}
\end{figure}

For a quantitative evaluation of the transition line, we require the  values of $E_0$ and $\Delta$ in both phases upon approaching the transition point. 
The ground-state energies are given by Eq.~(\ref{eLM}) and Eq.~(\ref{eMD}) for the LM and MD phase, respectively.  
We used exact diagonalization for a system of $N_s=30$ sites to extract an estimate for the excitation gaps. The excitation gap $\Delta$ as a function of the interaction ratio $J_2/J_1$ for different magnetic fields $h$ is shown in Fig.~\ref{fig:gaps}. 
We find that upon going from the magnetic field $h/J_1=2$ to $h/J_1=2.5$, the excitation gap in the LM phase becomes larger than that in the MD phase. We thus expect the bending of the first-order line to change upon increasing the magnetic field. In particular, for the case of $h/J_1=2.5$, considered already in Fig.~\ref{fig:h_25_nd}, the line bends slightly to the left. This is however hardly seen on the scale of the main panel of Fig.~\ref{fig:h_25_nd}. The bending is better seen in the inset, 
which also shows the location of the critical point as extracted from further QMC simulations (as detailed below). Note that based on the free-energy argument, we cannot determine the location of the critical point, but from the inset of  Fig.~\ref{fig:h_25_nd}, we find that its location roughly matches 
the estimated first-order transition line. 
The deviation that is visible in the inset can be explained as follows: the form Eq.~(\ref{eq:F_gap}) matches
the exact expression Eq.~(\ref{eq:MDthermo}) in the MD phase whereas in the LM
phase it neglects the dispersive nature of the excitations above the gap
$\Delta$. Consequently, the average excitation energy in the LM phase is
effectively larger than $\Delta$, such that the transition line
should indeed bend further towards smaller $J_2$.

\begin{figure}[t]
    \centering
    \includegraphics[width=0.99\columnwidth]{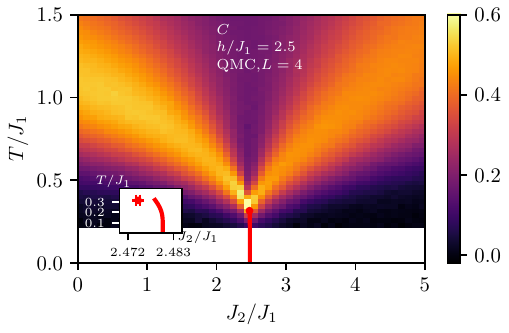}
    \caption{Specific heat in the vicinity of the LM-MD transition for $h/J_1=2.5$ as obtained from QMC for the system size with $L=4$. The red line shows the first-order transition line  obtained from comparing the free energies of both phases, extended up to the location of the critical point (symbol), as extracted from a finite-size analysis of the QMC data (see text for details). In the low-$T$ region (white), the QMC data exhibit large statistical fluctuations and have been cut off.}
    \label{fig:h_25_2dscan}
\end{figure}

The main panel of Fig.~\ref{fig:h_25_2dscan} shows the specific heat $C$ in the transition regime. Here, we observe two well pronounced lines of maxima that expand out from the location of the critical point, very similar to the behavior observed previously in related systems~\cite{Stapmanns2018,Jimenez2021,Weber2021}.

\begin{figure}[t]
    \centering
    \includegraphics[width=0.99\columnwidth]{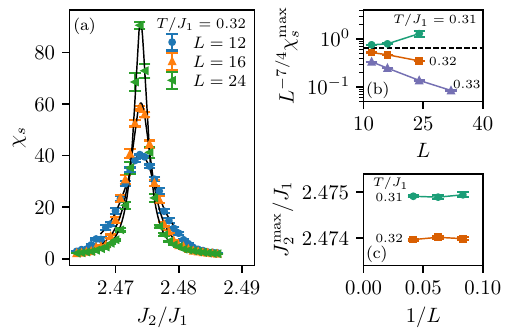}
    \caption{Left panel: Singlet susceptibility $\chi_s$ for a fixed temperature  $T/J_1=0.32$ across the transition region for $h/J_1=2.5$ and for various system sizes as obtained from QMC simulations. Right panel: finite-size scaling analysis of the peak value and its positions to extract the location of the critical point at $h/J_1=2.5$.}
    \label{fig:ising_point_h25}
\end{figure}

To accurately locate the critical point, we performed a finite-size scaling analysis for the fluctuations of the mean singlet occupancy. More specifically, we consider the corresponding singlet susceptibility~\cite{Stapmanns2018},
\begin{equation}
    \chi_s=\frac{\beta}{4N} \left( \left\langle \left(\sum_{d=1}^{2N} {n}_d\right)^2 \right\rangle
    -\left\langle \sum_{d=1}^{2N} {n}_d \right\rangle^2
    \right).
\end{equation}
In the left panel of Fig.~\ref{fig:ising_point_h25}, we show this quantity for different system sizes at a fixed temperature
$T/J_1=0.32$ across the transition region. The data exhibit pronounced maxima. Within the two-dimensional Ising universality
 of the critical point, the maximum value  scales as $\chi_s^\textrm{max}\propto L^{7/4}$ at
 criticality~\cite{Stapmanns2018}. This property can be used to extract the value of $T_c$ from performing a finite-size
 scaling analysis of the peak position, as shown in the upper right panel of Fig.~\ref{fig:ising_point_h25}, giving
 $T_c/J_1=0.315(5)$. From analyzing the corresponding values of $J_2/J_1$ of the peak position,  cf.\ the lower right panel of Fig.~\ref{fig:ising_point_h25}, we  can extract the critical coupling ratio $(J_2/J_1)_c=2.4745(5)$ as well. Together they give the estimated location of the critical point already shown in Figs.~\ref{fig:h_25_nd} and ~\ref{fig:h_25_2dscan}.

\begin{figure}[t]
    \centering
    \includegraphics[width=0.99\columnwidth]{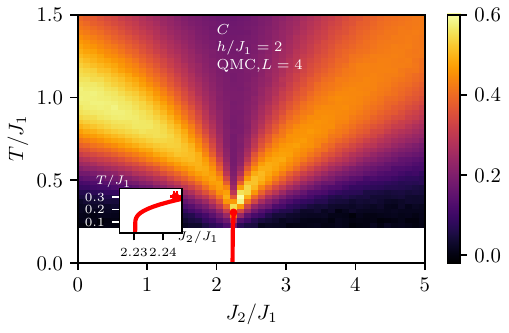}
    \caption{Specific heat in the vicinity of the LM-MD transition for $h/J_1=2$ as obtained from QMC for the system size with $L=4$. The red line shows the first-order transition line  obtained from comparing the free energies of both phases, extended up to the location of the critical point (symbol), as extracted from a finite-size analysis of the QMC data (see text for details). In the low-$T$ region (white), the QMC data exhibit large statistical fluctuations and have been cut off.}
    \label{fig:h_2_2dscan}
\end{figure}

We also performed a corresponding analysis at $h/J_1=2$. Here, according to the excitation gaps shown in
Fig.~\ref{fig:gaps}, we expect the first-order line to bend to the right instead. This is indeed confirmed by our analysis,
cf.\ the corresponding data for the specific heat shown in Fig.~\ref{fig:h_2_2dscan}.
The small deviations that one can see in the inset of Fig.~\ref{fig:h_2_2dscan}
can be explained by the same argument as in the case $h/J_1=2.5$, i.e.,
neglecting the dispersive nature of the excitations in the LM phase.

For the future, it would be interesting to  investigate the thermal properties of this model with respect to several other aspects, such as (i) how the KT transition lines of the SC phase merge with the discontinuous quantum phase transition between the SC and MD phases, and (ii) how the thermal properties of the DT phase can be quantitatively described by effective models of low-energy excitations, similar to the lattice-gas model for the MD phase. We hope that our investigations motivate further research on these challenging topics in the future. 

\section{Conclusions}\label{sec:conclusions}
In this article we considered the spin-1/2 Heisenberg antiferromagnet on the diamond-decorated square lattice in the presence of a finite magnetic field, using a combination of analytical arguments and  exact diagonalization, density matrix renormalization group, as well as sign-problem free stochastic series expansion quantum Monte Carlo simulations.

We identified the ground-state properties at finite magnetic field and  mention here  several aspects: (i) the previously identified zero-field Lieb-Mattis (LM), dimer-tetramer (DT) and monomer-dimer (MD) phases all extend to finite magnetic fields, with a magnetization 3/5, 0, and 1/5-plateau characterizing the LM, DT and MD regime, respectively, (ii) at intermediate fields, the DT phase vanishes and beyond this magnetic-field range, a direct discontinuous quantum phase transition takes place between the LM and MD phases, (iii) at high magnetic fields, in addition to the fully saturated paramagnetic phase (PM), a spin-canted (SC) phase with (quasi)-long-range order emerges. 

Additionally, we showed that in the MD regime for $J_2/J_1\geq4$, the low-temperature thermodynamic properties can be well described  in terms of  a simple effective lattice-gas model.

Motivated by related results in other quantum spin models, we showed that the direct, discontinuous quantum phase transition line between the LM and MD phase extends up to finite temperature, ending in a line of critical points that belong to the two-dimensional Ising universality class. We furthermore demonstrated that the slope of the transition line changes sign upon increasing the  strength of the magnetic field.

For the future it would certainly be interesting to extend this analysis to other phases such as the DT phase, and furthermore to investigate its excitations in more detail.

\acknowledgments

%
We acknowledge discussions with Lukas Weber, and support by the Deutsche Forschungsgemeinschaft (DFG) through Grant No.\ WE/3649/4-2 of the FOR 1807 and through RTG 1995, and thank the IT Center at RWTH Aachen University and  JSC Jülich for access to computing time through the JARA Center for Simulation and Data Science.
Part of the ED computations were carried out on the “osaka” cluster at the Centre de Calcul (CDC) of CY Cergy Paris Université.
We acknowledge support under the Stefanik program for Slovak-France bilateral projects SK-FR-19-0013 / 45125RC. KK and JS were financially supported under the grants VEGA 1/0105/20 and APVV-20-0150.  

\appendix
\newcommand{\tone}{\theta_{1}}
\newcommand{\td}{\theta_{d}}
\newcommand{\pone}{\phi_{1}}
\newcommand{\pd}{\phi_{d}}

\section{Ground states of the classical mixed spin-1 and spin-1/2 model on the Lieb lattice}\label{AppA}

\begin{figure}[t!]
    \centering
    \includegraphics[width=0.99\columnwidth]{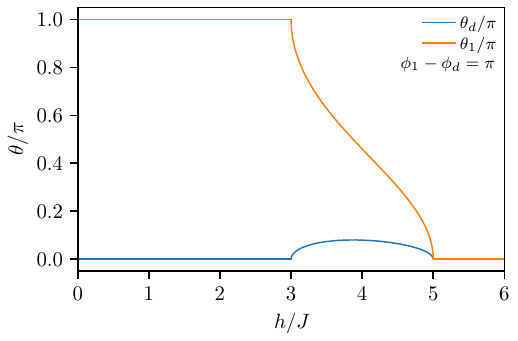}
    \caption{Angles $\theta/\pi$ of the spins of the classical mixed-spin Heisenberg model on the Lieb lattice as a function of the magnetic field $h/J$ at zero temperature.}
    \label{fig:classLieb}
\end{figure}
\begin{figure}[t]
    \centering
    \includegraphics[width=0.99\columnwidth]{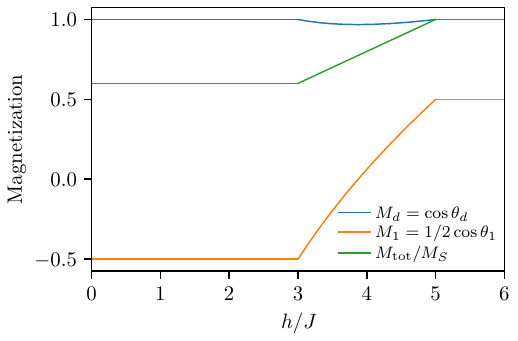}
    \caption{Local magnetizations $M_1$ and $M_d$ of the two inequivalent sites of the classical mixed-spin Heisenberg model on the Lieb lattice as well as the total magnetization $M_\mathrm{tot}$ (normalized by the saturated magnetization $M_S$) as a function of the magnetic field $h/J$ at zero temperature.}
    \label{fig:classicLieb_mag}
\end{figure}

In this appendix we consider a classical version of the  mixed spin-1 and spin-1/2 Heisenberg model in a magnetic field on the Lieb lattice. To this end we first note that the quantum Hamiltonian can be written as a sum of bond operators such that
\begin{align}
H = \sum \limits_{b=1}^{N_b} H_b= \sum \limits_{b=1}^{N_b} J\, \mathbf{S}_1\cdot \mathbf{S}_d - \frac{h}{4} S_1^z - \frac{h}{2}S_d^z , 
\end{align}
where $\mathbf{S}_1$ and $\mathbf{S}_d$ denote the spin-1/2 and spin-1 sites respectively and the sum runs over all bonds on the lattice. For a classical description, we replace the quantum spins by three-dimensional vectors of length 
 1/2 and 1, i.e., 
\begin{equation}
    \mathbf{S}_{1}\rightarrow \frac12
    \begin{pmatrix}
       \sin \tone \cos\pone\\
        \sin\tone \sin \pone\\
        \cos \tone
    \end{pmatrix}, \quad
    \mathbf{S}_d\rightarrow 
        \begin{pmatrix}
        \sin \td \cos\pd\\
        \sin\td \sin \pd\\
        \cos \td
    \end{pmatrix}.
\end{equation}
The bond terms in the classical model are then  given  by
\begin{align}
    H_{b} = &\frac{J}{2} \left(\sin \tone \sin\td \cos(\pone-\pd) + \cos \tone \cos\td \right)\nonumber \\ &-\frac{h}{2} \left(\frac14 \cos(\tone) + \cos(\td) \right) .
\end{align}
The ground state  of the total classical model is  obtained upon minimizing $H_b$ with respect to all four angles $\tone,\td,\pone,\pd$. Since $\theta_1$ and $\theta_d$ are restricted  between $0$ and $\pi$,  
minimizing with respect to both $\pd$ and $\pone$ yields $\pone-\pd=\pi$, i.e., an antiferromagnetic alignment of neighboring spins transverse to the field direction. This yields  
\begin{align}
    H_b = &\frac{J}{2} \left(-\sin \tone \sin\td + \cos \tone \cos\td \right)\nonumber \\ &-\frac{h}{2} \left(\frac14 \cos(\tone) + \cos(\td) \right) .
\end{align}
Differentiating with respect to $\tone$ and $\td$ gives the conditions
\begin{align}
    \frac{\partial H_b}{\partial \tone} &= \frac{J}{2}(-\cos \tone \sin \td - \sin \tone \cos \td ) + \frac{h}{8}\sin\tone=0 \label{eq:Hd1}\\
    \frac{\partial H_b}{\partial \td} &= \frac{J}{2} ( -\sin \tone \cos\td - \cos \tone \sin \td) + \frac{h}{2}\sin \td =0 \label{eq:Hdd} .
\end{align}
Subtracting Eq.~(\ref{eq:Hd1}) from Eq.~(\ref{eq:Hdd})  yields the relation $\sin \tone = 4 \sin \td$ which, when reinserted into Eq.~(\ref{eq:Hd1}), finally leads to 
\begin{align}
    \left(\frac{h}{8}-\frac{J}{8}\cos\tone - \frac{J}{2}\sqrt{1-\frac{1}{16}\sin^2\tone}\right)\sin\tone =0 . \label{eq:sol}
\end{align}
Based on Eq.~(\ref{eq:sol}), one can identify three different regimes, by requiring either factor to be zero, shown   in Fig.~\ref{fig:classLieb}. We identify first a ferrimagnetic (FI) regime for $0\leq h/J \leq {}3$, where the spins align in opposite directions with $\theta_1= \pi$ and $\theta_d=0$, with a ground-state energy of
\begin{align}
    E_\mathrm{FI}/N_b = -\frac{J}{2} - \frac{3}{8}h .
\end{align}
Next, we identify in the regime $3\leq h/J \leq 5$ a phase in which the spin directions change continuously  upon varying the magnetic field, given by
\begin{align}
    \cos\td {=} \frac {J}{8h} \left[ \left(\frac{h}{J}\right)^2 {+} 15 \right], \:
    \cos\tone {=} \frac {J}{2h} \left[ \left(\frac{h}{J}\right)^2 {-} 15 \right] .
\end{align}
Here, the spins are canted with respect to the direction of the magnetic field, forming biconical structures with the total magnetization 
$M_{tot}/M_s = (1/2 \cos \theta_1 +  2 \cos \theta_d) / (5/2) =  h/(5J)$.
Note that at small fields, $\mathbf{S}_1$ is  aligned in the opposite direction of the magnetic field. Upon increasing the magnetic-field strength, however, both spins align with the magnetic-field direction. Finally, we identify a fully saturated paramagnetic (PM) phase, where all spins align in direction of the magnetic field with $\theta_1=\theta_d=0$ and a ground-state energy
\begin{align}
    E_\mathrm{PM}/N_b = \frac{J}{2} - \frac{5}{8}h .
\end{align}

Figure~\ref{fig:classicLieb_mag} shows the local magnetization of both sites $\mathbf{S}_1$ and $\mathbf{S}_d$ as well as the (normalized) total magnetization as a function of the magnetic field. We note that this classical result is in good qualitative agreement with the results obtained for the quantum model, where the local magnetizations within the LM phase are slightly suppressed compared to their saturated values due to quantum fluctuations.

\section{Determination of the KT transition}\label{AppB}

\begin{figure}[t!]
    \centering
    \includegraphics[width=0.99\columnwidth]{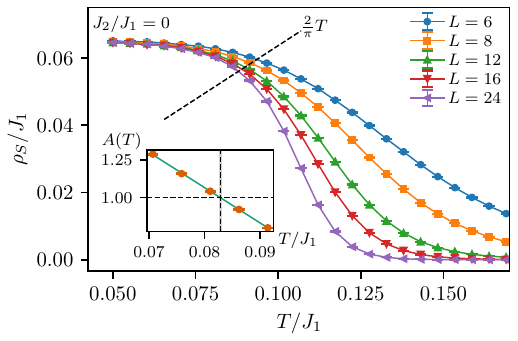}
    \caption{Spin stiffness $\rho_S$ for different system sizes $L$ as a
    function of temperature $T/J_1$ for the diamond-decorated square lattice model at $J_2/J_1=0$ and $h/J_1=4.5$. The dashed line denotes the scaling form of the universal jump. The inset shows the quantity $A(T)$ from the finite-size scaling analysis.  The KT transition temperature  is denoted by the dashed vertical line, where $A(T)=1$ holds, obtained using a linear fit (solid line).}
    \label{fig:appB}
\end{figure}

In this appendix, we detail the determination of the KT transition temperature $T_\mathrm{KT}$ within the SC phase. 
A standard means of identifying $T_\mathrm{KT}$ in $O(2)$-symmetric systems is based on the behavior of the spin stiffness $\rho_S$, which is  predicted to exhibit a universal jump of $\rho_S = 2\, T_\mathrm{KT}/\pi$ at $T_\mathrm{KT}$~\cite{Nelson1977}.
Within the SSE QMC approach, $\rho_S$ can be calculated from the spin winding number fluctuations \cite{Pollock1987,Sandvik1997,Caci2021}
\begin{equation}
    \rho_S = \frac{T}{2 A_\mathrm{uc}} \bigl(\langle W_x^2\rangle + \langle W_y^2\rangle\bigr),
\end{equation}
where $W_x$ and $W_y$ are the total winding numbers in the orthogonal $x$ and $y$ direction, respectively. Here $A_\mathrm{uc}$ is the unit cell area of the underlying Bravais lattice. For the diamond-decorated square lattice, $A_\mathrm{uc}=1$ holds.  
To extract $T_\mathrm{KT}$ from finite-size QMC data, we  follow the standard approach of Ref.~\cite{Harada1998}, which is based on the finite-size scaling form~\cite{Weber1988}
\begin{equation}
    \frac{\rho_S\,\pi}{2\,T} = A(T)\,\biggl( 1  + \frac{1}{2\,\log(L/L_0(T))}\biggr)
\end{equation}
that holds exactly at the transition point with $A(T_\mathrm{KT})=1$. We  fitted this finite-size dependence to the data for different temperatures, using $A(T)$ and $L_0(T)$ as fit parameters. This allows us to accurately estimate $T_\mathrm{KT}$, where $A(T_\mathrm{KT})=1$ holds. Our results from this approach are shown in Fig.~\ref{fig:appB}, and we obtain from this analysis an estimate of $T_\mathrm{KT}/J=0.08248(3)$ at $J_2/J_1=0$ and $h/J_1=4.5$. Performing the same analysis at different ratios of $J_2/J_1$ within the SC phase for $h/J_1=4.5$, we obtain similar values for the KT transition temperature. 


\bibliography{paper.bib}

\end{document}